\documentstyle[11pt,amsmath,amsfonts]{article}

\newcommand{\one}{\mbox{\tt 1}\hspace{-0.057 in}\mbox{\tt l}}

\def\Ket#1{|#1\rangle}

\newcommand{\x}{\mbox{\boldmath $x$}}
\newcommand{\xs}{\mbox{\boldmath $\scriptstyle x$}}
\newcommand{\y}{\mbox{\boldmath $y$}}
\newcommand{\ys}{\mbox{\boldmath $\scriptstyle y$}}
\newcommand{\z}{\mbox{\boldmath $z$}}
\newcommand{\zs}{\mbox{\boldmath $\scriptstyle z$}}
\newcommand{\gammas}{\mbox{$\scriptstyle \gamma$}}
\newcommand{\boldarrow}{\mbox{\boldmath $\swarrow$}}

\newcommand{\Tr}{\mbox{\rm\small Tr\ }}
\newcommand{\balpha}{\mbox{\boldmath $\alpha$}}
\newcommand{\balphas}{\mbox{\boldmath $\scriptstyle \alpha$}}
\newcommand{\bbeta}{\mbox{\boldmath $\beta$}}
\newcommand{\bbetas}{\mbox{\boldmath $\scriptstyle \beta$}}

\begin{document}

\title{Decoherence and linear entropy increase in the quantum baker's map}

\author{Andrei N. Soklakov and R\"udiger Schack\\
\\
{\it Department of Mathematics, 
Royal Holloway, University of London,}\\
{\it Egham, Surrey TW20 0EX, UK}}

\date{13 July 2001}

\maketitle

\begin{abstract}
  We show that the coarse-grained quantum baker's map exhibits a linear
  entropy increase at an asymptotic rate given by the Kolmogorov-Sinai entropy
  of the classical chaotic baker's map.  The starting point of our analysis is
  a symbolic representation of the map on a string of $N$ qubits, i.e., an
  $N$-bit register of a quantum computer.  To coarse-grain the quantum
  evolution, we make use of the decoherent histories formalism. As a
  byproduct, we show that the condition of medium decoherence holds
  asymptotically for the coarse-grained quantum baker's map.
\end{abstract}

The Kolmogorov-Sinai (KS) entropy of a classical dynamical system
\cite{Alekseev1981} quantifies the asymptotic rate at which information about
the initial conditions needs to be supplied in order to retain the ability to
predict the time-evolved system state with a fixed accuracy.  It can also be
viewed as the asymptotic linear rate of entropy increase of the coarse-grained
evolution of the dynamical system.  A positive KS entropy is one of the
simplest and most general criteria for classical chaos.  Several
generalizations of KS entropy to quantum mechanics have been proposed as
criteria for quantum chaos.
Refs.~\cite{Pechukas1982b,Zurek1994a,Alicki1994,Slomczynski1994} focus on
linear entropy increase, whereas
Refs.~\cite{Schack1993e,Schack1996b,Schack1996a} generalize the notion
of unpredictability, inherent in the concept of KS entropy, 
to quantum mechanics.

The dynamics of an isolated quantum system is unitary and therefore entropy
preserving.  The entropy can grow only if there is a source of
unpredictability such as coarse graining, measurement, or interaction with a
heat bath \cite{Caves1997b}. The same is true classically, where for example 
the entropy of a coarse-grained
probability distribution increases under  chaotic time evolution,
even though the Liouville equation preserves the entropy of the exact,
fine-grained distribution. 
Measurement as a source of unpredictability was
used in the definition of quantum dynamical
entropy~\cite{Slomczynski1994}, which has been conjectured to approach KS
entropy in the classical 
limit~\cite{Slomczynski1994,Slomczynski1998,Pakonski1999}.  
A linear growth of entropy for an inverted quantum harmonic
oscillator coupled to a heat bath has been established in
Ref.~\cite{Zurek1994a}.  Most results in this field are
obtained numerically (see, e.g.,
\cite{Alicki1996b,Kwapien1997,Habib1998,Miller1999}).
In this paper we derive rigorous results using coarse graining as
a source of unpredictability.

A systematic way to coarse-grain unitary quantum dynamics is provided by the
decoherent histories formalism
\cite{Griffiths1984,Omnes1988,Gell-Mann1990,Dowker1992}. In this formalism,
the quantum analogue of a coarse-graining of classical phase space takes the
form of a {\it coarse-grained history}.  The entropy of coarse-grained
histories has been defined and analyzed in
\cite{Gell-Mann1990,Hartle1998,Brun1999b}.  In this paper we give a rigorous
proof that the entropy of the coarse-grained quantum baker's map exhibits an
asymptotic linear growth of 1 bit per iteration, which equals the KS entropy
of the classical baker's map.  In order to prove this result, we first
establish that the coarse-grained histories satisfy the condition of
asymptotic medium decoherence \cite{Gell-Mann1990}.  Although the consistent
histories approach has been used before for the investigation of quantum
dissipative chaos \cite{Brun1995}, to our knowledge this is the first time
that the decoherence condition for histories has been rigorously established
for a chaotic quantum system.

The quantum baker's map \cite{Balazs1989,Saraceno1990} is a prototypical
quantum map invented for the theoretical investigation of quantum chaos.
During the last decade, it has been studied extensively (see, e.g.,
\cite{Soklakov2000a} and references therein). In this paper we consider a
class of quantum baker's maps defined in \cite{Schack2000a}. These maps admit
a symbolic description in terms of shifts on strings of qubits (two-state
systems) similar to 
classical symbolic dynamics \cite{Alekseev1981}.  They can also be
derived from the semiquantum maps introduced in~\cite{Saraceno1994a}. See
Ref.~\cite{Vallejos1999} for an application of symbolic methods to more
general maps. The formulation and proof of the theorems below is based on
the development of the symbolic description of the quantum baker's map
given in~\cite{Soklakov2000a,Soklakov2001a}.

Quantum baker's maps are defined on the $D$-dimensional Hilbert space
of the quantized unit square \cite{Weyl1950}. For consistency of units,
we let the quantum scale on ``phase space'' be $2\pi\hbar=1/D$.
Following Ref.~\cite{Saraceno1990}, we choose half-integer eigenvalues
$q_j=(j+{1\over2})/D$, $j=0,\ldots,D-1$, and $p_k=(k+{1\over2})/D$,
$k=0,\ldots,D-1$, of the discrete ``position'' and ``momentum''
operators $\hat q$ and $\hat p$, respectively, corresponding to
antiperiodic boundary conditions.  We further assume that $D=2^N$,
which is the dimension of the Hilbert space of $N$ qubits.

The $D=2^N$ dimensional Hilbert space modeling the unit square can be 
identified with the product space of $N$ qubits via
\begin{equation}
\Ket{q_j} =
\Ket{\xi_1}\otimes\Ket{\xi_2}\otimes\cdots\otimes\Ket{\xi_N}  \;,
\label{eq:tensor1}
\end{equation}
where $j=\sum_{l=1}^N \xi_l2^{N-l}$, $\xi_l\in\{0,1\}$,
and where each qubit has basis states $|0\rangle$ and $|1\rangle$.
We can write $q_j$ as a binary fraction, $q_j=0.\xi_1\xi_2\ldots\xi_N1$. 
We define the notation 
\begin{equation}
\Ket{.\xi_1\xi_2\ldots \xi_N} = e^{i\pi/2} \Ket{q_j} \;;
\label{eqtensor}
\end{equation}
see Ref.~\cite{Schack2000a} for the reason for the phase factor $e^{i\pi/2}$.
Momentum and position eigenstates are related through the quantum Fourier
transform operator $\hat F$ {\cite{Saraceno1990}}, i.e., 
$\hat F\Ket{q_k}=\Ket{p_k}$.

By applying the Fourier transform operator to the $n$ rightmost bits of
the position eigenstate $|.\xi_{n+1}\ldots\xi_N\xi_n\ldots\xi_1\rangle$, 
one obtains the family of states  \cite{Schack2000a}
\begin{eqnarray}                                                 \label{baker8}
|\xi_1\ldots\xi_n.\xi_{n+1}\ldots\xi_N\rangle&\equiv&
2^{-n/2}e^{i\pi(0.\xi_n\ldots\xi_11)}
|\xi_{n+1}\rangle
            \otimes\cdots\otimes
            |\xi_N\rangle \otimes\cr
    & & (|0\rangle+e^{2\pi i(0.\xi_11)}|1\rangle)
                     \otimes\cr
    & & (|0\rangle+e^{2\pi i(0.\xi_2\xi_11)}|1\rangle)
                      \otimes\cdots\otimes\cr
    & & (|0\rangle+e^{2\pi i(0.\xi_n\ldots\xi_11)}|1\rangle)\;,
\end{eqnarray}
where $1\leq n\leq N-1$.
For given $n$, these states form an orthonormal basis. 
The state (\ref{baker8}) is localized in both 
position and momentum: it is strictly localized within a position region 
of width $1/2^{N-n}$, centered at position 
$q=0.\xi_{n+1}\ldots\xi_N1$, and it is crudely localized within 
a momentum region of width $1/2^{n}$, centered at momentum 
$p=0.\xi_n\ldots\xi_11$.

For each $n$, $0\leq n\leq N-1$, a quantum baker's map can be defined by
\begin{equation}                                                 \label{baker9}
\hat{B}|\xi_1\ldots\xi_n.\xi_{n+1}\ldots\xi_N\rangle =
                    |\xi_1\ldots\xi_{n+1}.\xi_{n+2}\ldots\xi_N\rangle \;,
\end{equation}
where the dot is shifted by one position.  In phase-space language,
the map $\hat B$ takes a state localized at
$(q,p)=(0.\xi_{n+1}\ldots\xi_N1,0.\xi_n\ldots\xi_11)$ to a state localized at
$(q',p')=(0.\xi_{n+2}\ldots\xi_N1,0.\xi_{n+1}\ldots\xi_11)$, 
while it stretches the
state by a factor of two in the $q$ direction and squeezes it by a
factor of two in the $p$ direction.
For $n=N-1$, the map is the original quantum baker's map
as defined in Ref.~\cite{Saraceno1990}.

We are now in a position to introduce coarse-grained sets of histories. 
Let us first simplify our notation slightly. 
Given the dimensions $N$ and $n$, the dot in the definition (\ref{baker8}) is 
redundant. Thus, we will write from now on
\begin{equation} 
|\xi_1\ldots\xi_N\rangle\equiv|\xi_1\ldots\xi_n.\xi_{n+1}\ldots\xi_N\rangle\,,
\end{equation}
always keeping in mind the given values of $N$ and $n$. We introduce a set 
of projection operators,
\begin{equation}                                \label{coarseProjectors}
  {P}_{\ys}^{(l,r)}\equiv
\sum_{a_1,\ldots,a_l\atop b_1,\ldots,b_r}
|a_1\dots a_l\ \y\ b_1\dots b_r\rangle
\langle a_1\dots a_l\ \y\ b_1\dots b_r|\,,
\end{equation}
where the bold variable $\y$ denotes the binary string $\y=y_1\ldots
y_{N-l-r}$.  The operator ${P}_{\ys}^{(l,r)}$ is a projector on a
$2^{l+r}$-dimensional subspace labeled by the string $\y$. The $2^{N-l-r}$
projectors defined by all possible bit strings $\y$ form a complete set of
mutually orthogonal projectors, i.e., ${P}_{\ys}^{(l,r)}{P}_{\ys'}^{(l,r)}=0$
if $\y\ne\y'$ and $\sum_{\ys}{P}_{\ys}^{(l,r)}=\one$.  We can write each
${P}_{\ys}^{(l,r)}$ as a diagram
\begin{equation}                                           \label{Pdiagrams}
{P}_{\ys}^{(l,r)}\equiv\,(\,\underbrace{\Box\Box\dots\Box}_{l}
\ \y\ \underbrace{\Box\Box\dots \Box}_{r}\,)\;,
\end{equation}
where the empty boxes indicate $l$ leftmost and $r$ rightmost bits which are
coarse-grained over. For simplicity, we will always assume in the following 
that $l<n$ and $r<N-n$. In this case $l$ and $r$ acquire a more specific
meaning as the number of ``momentum'' and ``position'' bits 
ignored in the coarse-graining. 

For a given dynamics, a string of projectors defines a coarse-grained history.
 We define two types of 
histories, $h_{\vec{\ys}}$ and $h^c_{\ys}$. The history $h_{\vec{\ys}}$ is
defined as
\begin{eqnarray}                                      \label{Type3Histories}
h_{\vec{\ys}}&\equiv&
                ({P}^{(l,r)}_{\ys^1},{P}^{(l,r)}_{\ys^2},
\dots,{P}^{(l,r)}_{\ys^k})_{\phantom{|_{|_{|_{|_|}}}}} \cr
&=&
\big( \ 
                         \underbrace{\Box\Box\dots\Box}_{l}
            \ \y^1\ \underbrace{\Box\Box\dots \Box}_{r}\ , \cr
                              && 
 \phantom{\big( \ }
                         \underbrace{\Box\Box\dots\Box}_{l}
            \ \y^2\ \underbrace{\Box\Box\dots \Box}_{r}\ ,\; \dots\;,   \cr
            &&    \phantom{|}        \cr
                              &&
 \phantom{\big( \ }
                        \underbrace{\Box\Box\dots\Box}_{l}
            \ \y^k\ \underbrace{\Box\Box\dots \Box}_{r} \   \big{)}\;,
\end{eqnarray}
where $\vec{\y}=(\y^1,\ldots,\y^k)$. Since for
each $t=1,\ldots,k$, the projectors ${P}^{(l,r)}_{\ys^t}$ form a
complete set of  mutually 
orthogonal projectors, the histories $\{ h_{\vec{\ys}} \}$ are said
to form an exhaustive set of mutually exclusive histories. They are a special
case of the more general sets of histories introduced in
Refs.~\cite{Griffiths1984,Omnes1988,Gell-Mann1990}.

The second type of histories considered here is defined by a further coarse-graining of 
the histories $\{ h_{\vec{\ys}} \}$, consisting of a summation over the first $k-1$ projectors
in~(\ref{Type3Histories}):
\begin{equation}                              
h^c_{\ys}\equiv
           (\underbrace{\one,\dots,\one}_{k-1{\rm\ times}},{P}^{(l,r)}_{\ys})\;.
\end{equation}
The histories $\{ h_{\ys}^{c} \}$
 also form an exhaustive set of mutually exclusive histories. 

Starting from some initial state $\rho_0$,
the coarse-grained evolution of the quantum baker's map
$\hat B$
is characterized by a {\it decoherence functional}. For the histories
$\{h_{\vec{\ys}} \}$, the decoherence functional is given by     
\begin{equation}                                      \label{dfunc}  
{\mathcal D}[\rho_0, \,h_{\vec{\ys}},
                                    h_{\vec{\zs}}   ]
=
                \Tr[P^{(l,r)}_{\ys^k}\hat{B} P^{(l,r)}_{\ys^{k-1}}
                \hat{B}\cdots P^{(l,r)}_{\ys^1}
\hat{B}\rho_0 \hat{B}^\dag P^{(l,r)}_{\zs^1}
   \cdots
         \hat{B}^\dag P^{(l,r)}_{\zs^{k-1}}\hat{B}^\dag P^{(l,r)}_{\zs^k}] \;,
\end{equation}
and for the  histories $\{ h_{\ys}^{c} \}$, by
\begin{equation}                                           \label{coarse-dfunc}
{\cal D}[\rho_0,h^c_{\ys}, h^c_{\zs}]=
  \Tr[{P}^{(l,r)}_{\ys}\hat{B}^k\rho_0(\hat{B}^\dag)^k{P}^{(l,r)}_{\zs} ]\,.
\end{equation}
In both cases, the number of iterations of the map, $k$, 
is assumed to satisfy the 
inequality $k<r$. In the following we assume
that the initial state is proportional to one of the projectors defined in 
Eq.~(\ref{coarseProjectors}), i.e.,
\begin{eqnarray}                                   \label{InitialState}
\rho_0=\rho^{{\scriptscriptstyle{(l,r)}}}_{\xs}
&\equiv&2^{-(l+r)}{P}^{(l,r)}_{\xs} \cr
&=&2^{-(l+r)}\,(\,\underbrace{\Box\Box\dots\Box}_{l}
\ \x\ \underbrace{\Box\Box\dots \Box}_{r}\,)\;.
\end{eqnarray}

If the off-diagonal elements of the decoherence functional vanish, the set
of histories is said to be decoherent (more precisely, this is the condition 
of medium decoherence \cite{Gell-Mann1990}). 
It follows directly from the cyclic property
of the trace that 
the coarse decoherence functional (\ref{coarse-dfunc})
satisfies the decoherence condition:
\begin{equation}                                           \label{coarse-decoherence}
{\cal D}[\rho_0,h^c_{\ys}, h^c_{\zs}] = 0 \;\; 
  \mbox{\rm if} \;\; \y\ne\z \;.
\end{equation}
For its diagonal elements, we have

\noindent{\bf Theorem 1:}
{\it Fix two strings $\x$ and $\y$ of the same length:
$|\x|=|\y|=c$. For any two strings $\balpha$ and $\bbeta$
such  that $|\balpha|=|\bbeta|=k$, where $k$ is a fixed
number of iterations, $k<r$, we have
\begin{equation}                                           \label{rGREATERka}
{\cal D}[\rho^{{\scriptscriptstyle{(l,r)}}}_{\balphas\xs},
h^c_{\ys\bbetas}, h^c_{\ys\bbetas}]
=2^{-k}\delta{\,}_{\xs}^{\ys}-O(\frac{l+r}{2^{l-k}})\,,
\end{equation}
where
$\balpha\x$ denotes concatenation of the strings $\balpha$ and $\x$ and 
similarly for $\y\bbeta$, and where $\delta{\,}_{\xs}^{\ys}$ denotes
the Kronecker delta function.}
 The proof of this and the results below
will be given in a subsequent longer paper.

Since the decoherence condition is satisfied, we can interpret the diagonal
elements~(\ref{rGREATERka}) as probabilities. We see that there is no
single dominant history. Instead, after the $k$-th step there are $2^k$
different histories each having asymptotically the same probability,
$2^{-k}$. These histories are defined by the condition
$\x=\y$, i.e., a shift of $k$ binary positions to the left:
\begin{eqnarray} 
\underbrace{\Box\Box\dots\Box}_{l}\;
                     \ \alpha_1
                                       \dots\alpha_k\;
                  \underline{\x_1\dots \x_c} \;
   \underbrace{\Box\Box\dots \Box}_{r}\ , \cr
   \swarrow_{\phantom{|_{|_{|_|}}}}\ \ \ \ \ \ \ \ \ \ \ \ \ \ \ \ \ \ \ \ \ \ \cr
 \underbrace{\Box\Box\dots\Box}_{l}\;
             \overline{ \ \y_1 \dots \y_c }   \;
                     \beta_1
                                  \dots\beta_k\;
   \underbrace{\Box\Box\dots \Box}_{r}\;.
\end{eqnarray}
During this transformation
the bits of $\balpha$ are lost as they reach the scale
at which the momentum becomes coarse-grained.
At the same time $k$ unspecified (i.e., random)
position bits $\beta_1\dots\beta_k$ enter the relevant section
of the string.  At each step the number of histories with
significant probability doubles, as each history branches
into two equiprobable histories.
This means there is a loss of one bit of information per iteration.

We now give a precise formulation of this information loss.
Since the set
of histories $\{ h^c_{\ys} \}$ is decoherent, we can define its entropy
\cite{Brun1999b,Hartle1998,Gell-Mann1990},
\begin{equation}
H(\{h^c_{\ys}\}) 
\equiv   -\sum_{\ys }p(h^c_{\ys})\log_2p(h^c_{\ys})\,,
\end{equation}
where 
$p(h^c_{\ys}) = {\cal D}[\rho^{{\scriptscriptstyle{(l,r)}}}_{\xs},
h^c_{\ys}, h^c_{\ys}]$.
Using theorem 1, we find that
\begin{equation}    \label{coarseentropy}
H(\{h^c_{\ys}\}) = k+
O(\frac{(l+r)\log_2(l+r)}{2^{l-k}})\,.
\end{equation}

The results for the very coarse histories $h^c_{\ys}$ depend in part on the
fact that the decoherence condition is trivially satisfied for these
histories. In the more interesting case of the less coarse-grained 
histories $\{h_{\vec{\ys}}\}$, 
the decoherence condition is satisfied only asymptotically.
The following theorem establishes
this asymptotic decoherence and gives asymptotic values
for the diagonal elements of the decoherence functional. \\
{\bf Theorem 2:}  {\it
Fix any integer $\gamma\geq 1$, any string $\x$ of length $|\x|=\gamma$,
and any two ordered sequences of strings 
$\vec{\y}=(\y^1,\y^2,\dots,\y^k)$
and
$\vec{\z}=(\z^1,\z^2,\dots,\z^k)$
such that $|\y^j|=|\z^j|=\gamma$, $j=1,\dots,k$, where $k$ is the
number of iterations, $k<r$. For sufficiently large $l$ we have then:
\begin{align}                                                            
{\mathcal D}[\rho^{\scriptscriptstyle (l,r)}_{\xs}, \,h_{\vec{\ys}},
                                    h_{\vec{\zs}}   ]
&=
2^{-k}
\left(\prod_{j=1}^{k}\delta_{\ys^j}^{\zs^j}
\right)
\left(
\prod_{j=1}^{k-1}
\delta{}_{\ys^{j+1}_{1:\gammas-1}}^{\;\ys^{j}_{2:\gammas}}
\delta{}_{\ys^j_{1}}^{\xs_{j+1}}
\right)
\delta{}_{\ys^k_{1:\gamma-k}}^{\xs_{k+1:\gamma}}
       +\,O(\frac{l+r-k}{2^{l-2(k^2+k)}})            \nonumber
\end{align}
\begin{align}
{\phantom{{\mathcal D}}}
&=2^{-k}
\underbrace{
           \left(\prod_{j=1}^{k}\delta_{\ys^j}^{\zs^j}\right)
}_{{\rm diagonal}}
\cdot
\underbrace{\left(
\delta{}_{\ys^1_{1:\gamma-1}}^{\xs_{2:\gamma}}
\prod_{j=1}^{k-1}
\delta{}_{\ys^{j+1}_{1:\gammas-1}}^{\;\ys^{j}_{2:\gammas}}
\right)}_{{\rm step-by-step\ shift}}
\cdot
\underbrace{ \Bigg{(}
\delta{}_{\ys^k_{1:\gamma-k}}^{\xs_{k+1:\gamma}}
\Bigg{)} }_{k{\rm th\ shift }} 
\;\; +\,O(\frac{l+r-k}{2^{l-2(k^2+k)}})\;.
\end{align}
The second equality provides a somewhat
redundant but more transparent formulation of the theorem.}

We see that the expression in the first parentheses is zero for all
off-diagonal elements of the decoherence functional.
This implies that in the limit of large $l$ all off-diagonal elements of the
decoherence functional vanish, which establishes the medium decoherence
condition. The diagonal elements of the decoherence functional
can therefore be interpreted as probabilities of the corresponding
histories (see Ref.~\cite{Dowker1992} for a discussion of approximate
decoherence). Asymptotically, only $2^{k}$ diagonal elements 
are nonzero. Moreover, the error terms are exponentially small.
As in the case of the coarse histories considered above, 
there are $2^{k}$ histories with asymptotically
equal probabilities. The number of such
histories doubles after each step resulting in a loss of information
at the rate of 1 bit per step. The conditions satisfied by the histories with
nonzero probabilities are also similar to the previous case. Here,
each of these
histories is a sequence of $k$ projectors and each of those
projectors is related to the initial state via a shift according to
the position of the projector in the history:
\begin{eqnarray}
& \underbrace{\Box\Box\dots\Box}_{l}\;
                     \ \x_1
                  \underline{\x_2\dots\x_{\gamma-2}\x_{\gamma-1} \x_\gamma } \;
   \underbrace{\Box\Box\dots \Box}_{r}\ , \cr
& \ \ \ \ \ \ \boldarrow_{\phantom{|_{|_{|_|}}}} \cr
& \underbrace{\Box\Box\dots\Box}_{l}\;
             \overline{ \ \y^1_1\underline{\y^1_2\dots\y^1_{\gamma-2} \y^1_{\gamma-1}}}
                     \underline{\; y^1_\gamma}\;
   \underbrace{\Box\Box\dots \Box}_{r}\ , \cr
& \ \ \ \ \ \ \boldarrow_{\phantom{|_{|_{|_|}}}} \cr
& \underbrace{\Box\Box\dots\Box}_{l}\;
             \overline{ \ \y^2_1 \underline{\y^2_2\dots\,\y^2_{\gamma-2} y^2_{\gamma-1}}}
                     \underline{\; y^2_\gamma}\;
   \underbrace{\Box\Box\dots \Box}_{r}\ , \cr
& \ \ \ \ \boldarrow \cr
& \ \ \ \ \dots \cr
& \ \ \ \ \ \ \boldarrow_{\phantom{|_{|_{|_|}}}} \cr
& \underbrace{\Box\Box\dots\Box}_{l}\;
             \overline{ \y^k_1 \dots\y^k_{\gamma-k} y^k_{\gamma-k+1}
                       \!\dots}\, y^k_\gamma\,
   \underbrace{\Box\Box\dots \Box}_{r}\ .
\end{eqnarray}
In this diagram the first line represents the initial condition
$\rho^{{\scriptscriptstyle{(l,r)}}}_{\xs}$. The subsequent lines
correspond to the projectors $P^{(l,r)}_{\ys^1},\dots,P^{(l,r)}_{\ys^k}$
in the history. The bold face is used to indicate the bits which are
completely determined by the initial condition for those histories with
asymptotically nonzero probability. Such histories satisfy the
step-by-step shift condition denoted on the diagram by the arrows
and lines: for example, the substring $\x_2\dots\x_\gamma$ is shifted onto
the substring $\y^1_1\dots\y^1_{\gamma-1}$.
For the entire history, therefore,
there are only $k$ independent bits which can be chosen
arbitrarily, given the step-by-step shift constraint. We recover the
coarse-histories case considered above if we
choose $y^k_{\gamma-k+1}\dots y^k_\gamma$ as independent and record only
the very last projector, ignoring the rest of the trajectory.

The entropy of the approximately decoherent set
of histories $\{h_{\vec{\ys}}\}$ is
\begin{equation}
H(\{h_{\vec{\ys}} \}) 
 = -\sum_{\vec{\ys}}p(h_{\vec{\ys}})\log_2p(h_{\vec{\ys}})\,,
\end{equation}
where $p(h_{\vec{\ys}}) = 
{\mathcal D}[\rho^{\scriptscriptstyle (l,r)}_{\xs}, \,h_{\vec{\ys}},
                                    h_{\vec{\ys}}]$.
It follows then from theorem 2 that 
\begin{equation}               \label{main}
H(\{h_{\vec{\ys}} \}) = k+
O(\frac{(l+r-k)\log_2(l+r-k)}{2^{l-2(k^2+k)}})\,.
\end{equation}
In the limit of large $l$, for any fixed number of iterations, $k$, the
entropy of the coarse-grained quantum baker's map approaches the value of $k$
bits, i.e., 1 bit per iteration, which is the KS entropy 
of the classical baker's map.  Due to
the $k^2$ term in the denominator, the bound on the error term is not as tight
as in Eq.~(\ref{coarseentropy}). We believe that this bound can be further
improved.\\


\end{document}